\documentclass[aps, pra, reprint, floatfix, superscriptaddress]{revtex4-2}
\usepackage{amsmath}
\usepackage{physics}
\usepackage{siunitx}
\usepackage{hyperref}
\usepackage{xcolor}
\usepackage[version=4]{mhchem}
\usepackage{graphicx}

\usepackage{pdfpages} 
\makeatletter
\AtBeginDocument{\let\LS@rot\@undefined}
\makeatother

\usepackage[bbgreekl]{mathbbol}
\DeclareSymbolFontAlphabet{\mathbbm}{bbold}
\usepackage{enumerate}

\newcommand{\suppDqmc}{I}
\newcommand{\suppComp}{II}
\newcommand{\suppRcdw}{III}
\newcommand{\suppDiff}{IV}
\newcommand{\suppFSS}{V}
\newcommand{\suppErr}{VI}
\newcommand{\suppChi}{VII}
\newcommand{\suppDen}{VIII}
\newcommand{\suppSym}{IX}

\begin{document}

\title{Trion formation and ordering in the attractive SU(3) Fermi-Hubbard model}
\author{Jonathan Stepp}
\email{jds28@rice.edu}
\affiliation{Department of Physics and Astronomy, Rice University, Houston, Texas 77005, USA}
\author{Eduardo Ibarra-Garc{\'i}a-Padilla}
\affiliation{Department of Physics, Harvey Mudd College, 301 Platt Blvd, Claremont, CA 91711, USA}
\author{Richard T. Scalettar}
\affiliation{Department of Physics, University of California, Davis, California 95616, USA}
\author{Kaden R. A. Hazzard}
\affiliation{Department of Physics and Astronomy, Rice University, Houston, Texas 77005, USA}
\affiliation{Smalley-Curl Institute, Houston, Texas 77005, USA}

\date{May 6, 2026}

\begin{abstract}
Recent advances in microwave shielding have increased the stability and control of large numbers of polar molecules, allowing for the first realization of a molecular Bose-Einstein condensate.
Remarkably, it was also recently realized that shielded polar molecules exhibit an SU($N$) symmetry among their hyperfine states, opening the door to SU($N$) systems with larger $N$, bosonic particle statistics, and tunable interactions --- both repulsive and attractive.
Motivated by these results, we have studied the SU(3) attractive Fermi-Hubbard model (FHM) on a square lattice.
Using the Determinant Quantum Monte Carlo (DQMC) method, we explore the finite-temperature phase diagram and provide evidence for three distinct regions --- a three-component Fermi liquid (FL) region, a ``trion'' liquid (TL) region, and an ordered Charge Density Wave (CDW) phase.
The CDW phase is stable at finite temperature (in contrast to the SU(2) CDW), while the FL to TL crossover appears to point to a quantum phase transition at zero temperature.
Our method extends straightforwardly to larger \( N \) and is sign-problem free for even values of \( N \).
With these results, we demonstrate the potential physics enabled by using polar molecules as a quantum simulation platform for the attractive SU($N$) FHM.
\end{abstract}

\maketitle

\textit{Introduction ---}
The last few decades have seen a resurgence of interest in SU($N$)-symmetric extensions of interacting-fermion lattice models, such as the \( t \)-\( J \) model, Heisenberg model, and Fermi-Hubbard model (FHM) \cite{affleck1988,read1989,sachdev1990}.
These SU($N$) models have been studied to approximate ultracold atoms with several hyperfine states \cite{honerkamp2004a, honerkamp2004}, where the SU($N$) symmetry was seen more as a loose approximation.
Then, the discovery of a precise SU($N$) symmetry in the nuclear spin of alkaline-earth atoms (AEAs) \cite{wu2003,gorshkov2010,cazalilla2014,he2019} sparked detailed theoretical and numerical work to study phases and the equation of state in these newly accessible models.
In recent years, multiple groups have experimentally realized SU($N$) lattice models \cite{taie2012,hofrichter2016,pasqualetti2024,tusi2022,taie2022} and many numerical methods and theoretical approaches have been developed to study them \cite{hermele2009,wang2014,zhou2016,xu2018,unukovych2021,ibarra-garcia-padilla2021,singh2022,ibarra-garcia-padilla2023,feng2023,botzung2024,kozik2024,schlomer2024,ibarra-garcia-padilla2024}.

However, with these AEAs come inherent limitations. Most relevant to the present work, interactions between AEAs cannot be controlled by a magnetic Feshbach resonance and so are always repulsive (i.e. have a positive \( s \)-wave scattering length).
The attractive case has been studied using other atoms --- such as three hyperfine levels of \ce{^6 Li} \cite{ottenstein2008} --- but tuning the interactions to the SU(3)-symmetric point has been a persistent challenge.
So, while the attractive SU($N$) FHM has been studied as a toy model of a wide array of phenomena --- from baryon formation in QCD \cite{rapp2008,xu2023} to charge-\(4e\) superconductivity \cite{lecheminant2005,capponi2007,capponi2008,soldini2024} --- there has not been an experimental platform to realize this model and theoretical investigations have mainly focused on the repulsive case.

Additionally, attractive SU($N$) models are a natural stage for studying many-body bound states.
In the continuum, the formation of three-particle ``trions'' has been studied along with the infinite ladder of Efimov states \cite{naidon2017,floerchinger2009,nishida2012}.
In lattice systems, few-body exact-diagonalization has shown the existence of both on-site and off-site trions and studied the critical interaction strength for trion formation in one and two dimensions \cite{pohlmann2013}.
The attractive SU($N$) FHM offers an opportunity to study how the formation of composite particles affects the many-body physics of a system.
Early mean field theory and renormalization group (RG) work focused on a quantum phase transition (QPT) from a trion phase to a many-body generalization of a Bardeen-Cooper-Schrieffer (BCS) state called a ``color superfluid'' (CSF) \cite{honerkamp2004,honerkamp2004a,rapp2007,rapp2008}.
There are many open questions concerning the finite temperature behavior of these phases and other ordered phases for the attractive SU($N$) FHM, especially in two dimensions.

Recently, it has become clear that ultracold molecule experiments employing electromagnetic shielding techniques provide an exciting new SU($N$) system, which may be able to experimentally realize the attractive SU($N$) FHM.
Cooling molecules down to degenerate quantum gases has historically been challenging due to large two-body loss rates that impair evaporative cooling.
Static electric field shielding has enabled the cooling of a degenerate Fermi gas of molecules \cite{valtolina2020}, and more recently, microwave shielding \cite{karman2018} has enabled even further cooling below the Fermi temperature \cite{schindewolf2022} and the first observation of a dipolar molecular BEC \cite{bigagli2024}.
Importantly for our purposes, this microwave shielding causes an effective SU($N$) symmetry \cite{mukherjee2025,mukherjee2025a}.
The interaction strength can be tuned across a wide range and can be either repulsive or attractive.
So, amidst the many proposals for using shielded dipolar molecules to study new physics \cite{cornish2024}, we believe these molecules will be useful for answering long-held questions about the attractive SU($N$) FHM.

In this Letter, we employ a Determinant Quantum Monte Carlo (DQMC) method designed for the attractive SU($N$) FHM to explore the phase diagram of the \( N=3 \) case on a square lattice.
We find evidence of two distinct phase transitions --- a ``trion formation'' QPT and an ``ordering'' transition into a charge density wave (CDW) phase, which persists to finite temperature.
We present evidence for these two transitions and develop approximations that are valid deep within the different phases.
We also compare our results to previous theoretical work on the zero-temperature phase diagram of this model and discuss the viability of observing these phases in an ultracold molecule experiment.

\textit{Model and Numerical Methods ---}
The attractive, SU(3)-symmetric FHM is given by the Hamiltonian
\begin{align}
   H &= \overbrace{-t \sum_{\sigma, \langle i,j \rangle}^{} \Big[ c^{\dagger}_{i\sigma} c^{\phantom \dagger}_{j\sigma} + \text{h.c.}\Big]}^{K} - \mu \sum_{\sigma,i}^{} n_{i\sigma} \nonumber \\
   &\quad  \underbrace{-\frac{U}{2}\sum_{i} \left(\sum_{\sigma}^{} n_{i\sigma} - \frac{3}{2}\right)^2}_{V},  \label{eqn:ham}
\end{align}
where \(c_{i\sigma}^\dagger, c_{i\sigma}^{\phantom \dagger}\) are the creation and annihilation operators for a molecule in spin state \(\sigma\) on lattice site \(i\), \(n_{i\sigma} = c^\dagger_{i\sigma} c^{\phantom \dagger}_{i\sigma}\), \(\langle i, j \rangle\) denotes a sum over nearest-neighbor pairs, \(t\) is the tunneling energy, \(U\) is the on-site attraction, \(\mu\) is the chemical potential, $K$ is the kinetic energy operator and $V$ is the potential energy operator.
Since we study the \( N=3 \) case, \( \sigma \) runs over three molecular spin states.
Here, \(\mu = 0\) corresponds to half-filling (\( n_\sigma = 1/2 \)).

To study this model at finite temperature, we use a DQMC method \cite{blankenbecler1981,scalapino1981} formulated for the attractive SU($N$) FHM as detailed in Supplemental Material \cite{supp}, Sec. \suppDqmc.
We compute finite-temperature expectation values such as \( \langle n_{i\sigma}\rangle \), \( \langle n_{i\sigma} n_{j\tau} \rangle \), and \( \langle n_{i\sigma} n_{i\tau} n_{i\nu} \rangle \) for molecular spin states \( \sigma,\tau,\nu \) and site indices \( i,j \) on square lattices with side length $L$ over a range of \( U/t \), \( \mu/t \), and \( T/t \), where $T$ is the temperature and we have set $k_B = 1$.
Quantities which require differentiation with respect to $\mu$ or $T$ were calculated by comparing neighboring data points.
All data shown is for \( L = 10 \) unless specified to be a finite-size extrapolation.
The DQMC algorithm works by discretizing the inverse temperature into $M$ imaginary time steps, where $1/T = M \Delta \tau$.
The details of the differentiation process, finite-size extrapolation, a comparison of error sources and plots with representative data and density plots are given in the Supplemental Material \cite{supp} Secs. \suppDiff, \suppFSS, \suppErr, and \suppDen respectively. 

\begin{figure}[t]
   \includegraphics[width=\columnwidth]{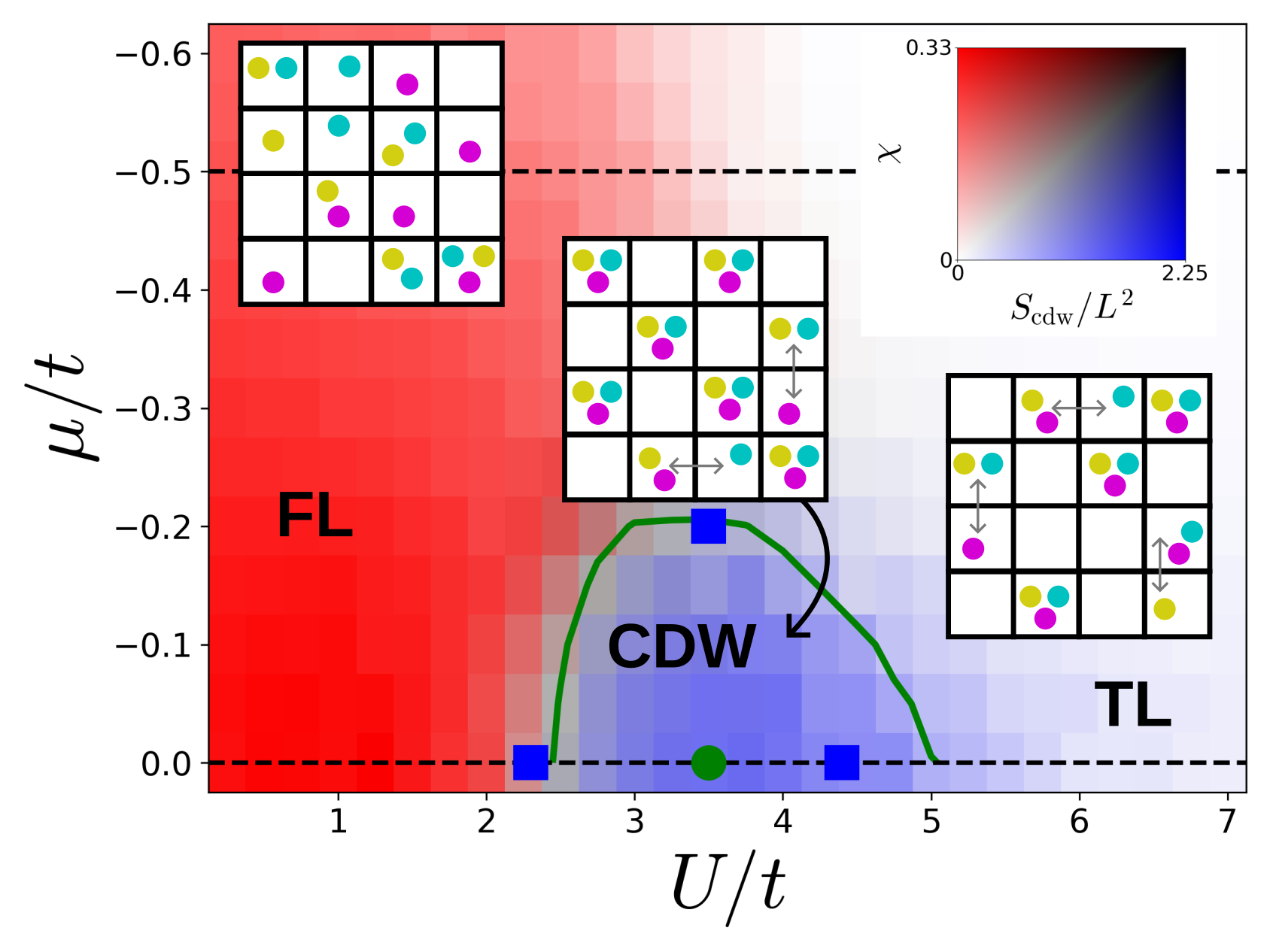}
   \caption{\label{fig:phase_diagram} Finite temperature phase diagram at \( T = t/3 \) with representative configurations for the three zero-temperature phases: the Fermi liquid of individual molecules (FL), trion Fermi liquid of bound triples of molecules (TL) and Charge Density Wave (CDW) phase. The inset shows a legend for the structure factor, \(S_{\rm cdw}/L^2 \), and difference susceptibility, \(\chi\). Dashed lines indicate the cuts taken in Fig.~\ref{fig:mu_cut}. The green dot shows the \( (U, \mu) \) point where the data in Fig.~\ref{fig:temp_cut} was taken. The green contour line is a guide for the eye while the blue squares mark the critical values as indicated by the invariant correlation ratio.}
\end{figure}

\begin{figure}[t]
   \includegraphics[width=\columnwidth]{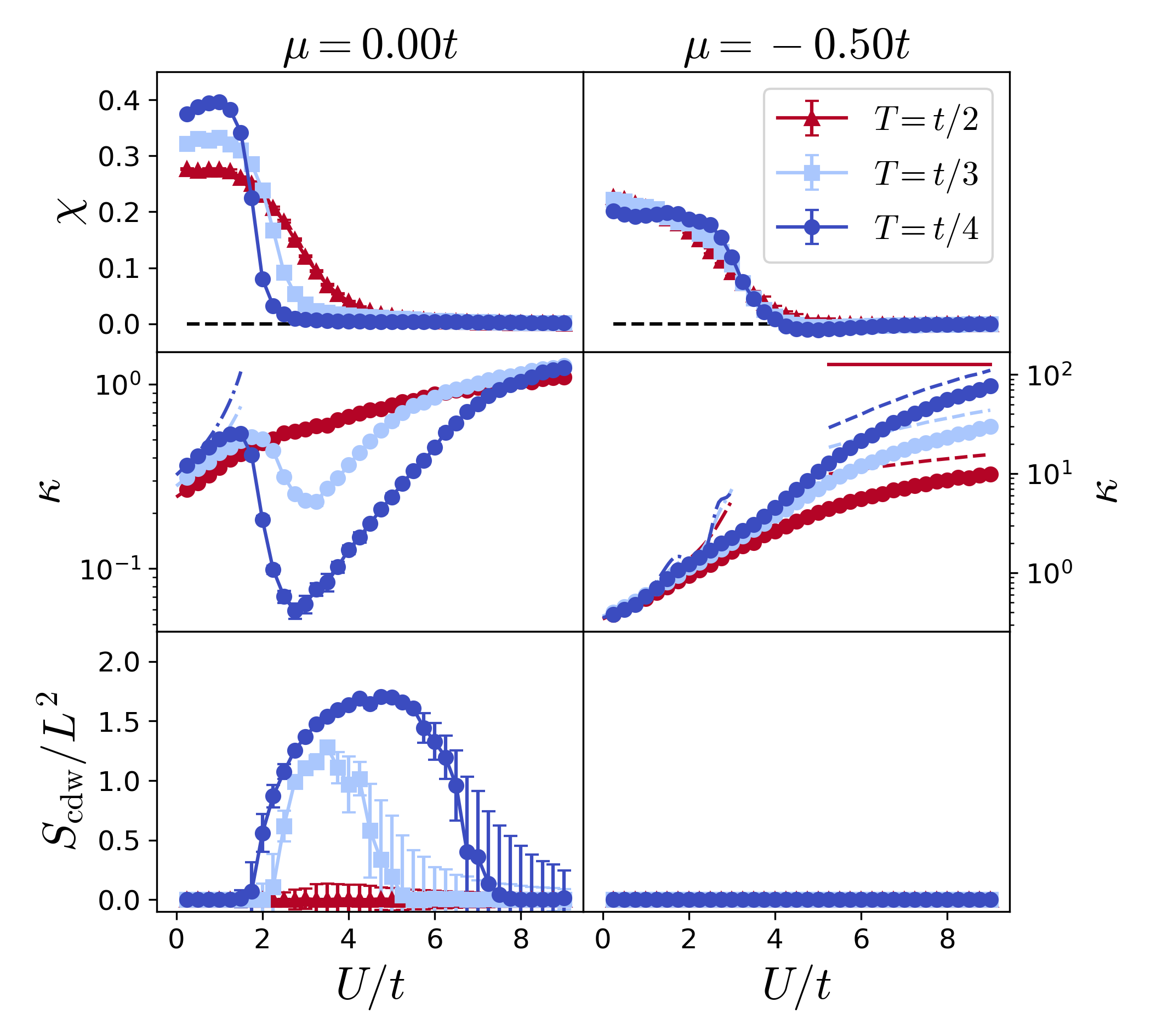}
   \caption{\label{fig:mu_cut} Difference susceptibility, isothermal compressibility, and structure factor (finite-size extrapolation) as a function of \( U/t \) for three temperatures.
     Left: \( \mu = 0 \), Right \( \mu = -0.5 t \).
     The dashed lines for large \( U \) in the \( \kappa \) plot indicate the classical ideal trion gas approximation, the dash-dot lines on the low-\( U \) side show the weakly-interacting mean field approximation and the solid line shows the atomic limit \(\kappa\) at \( T = t/2 \). Error bars in the $\kappa$ and $\chi$ plots show statistical error, while error bars in the $S_{\rm cdw}$ plot show the difference between the finite-size extrapolation and the $L=10$ data.}
\end{figure}

\begin{figure}
   \includegraphics[width=\columnwidth]{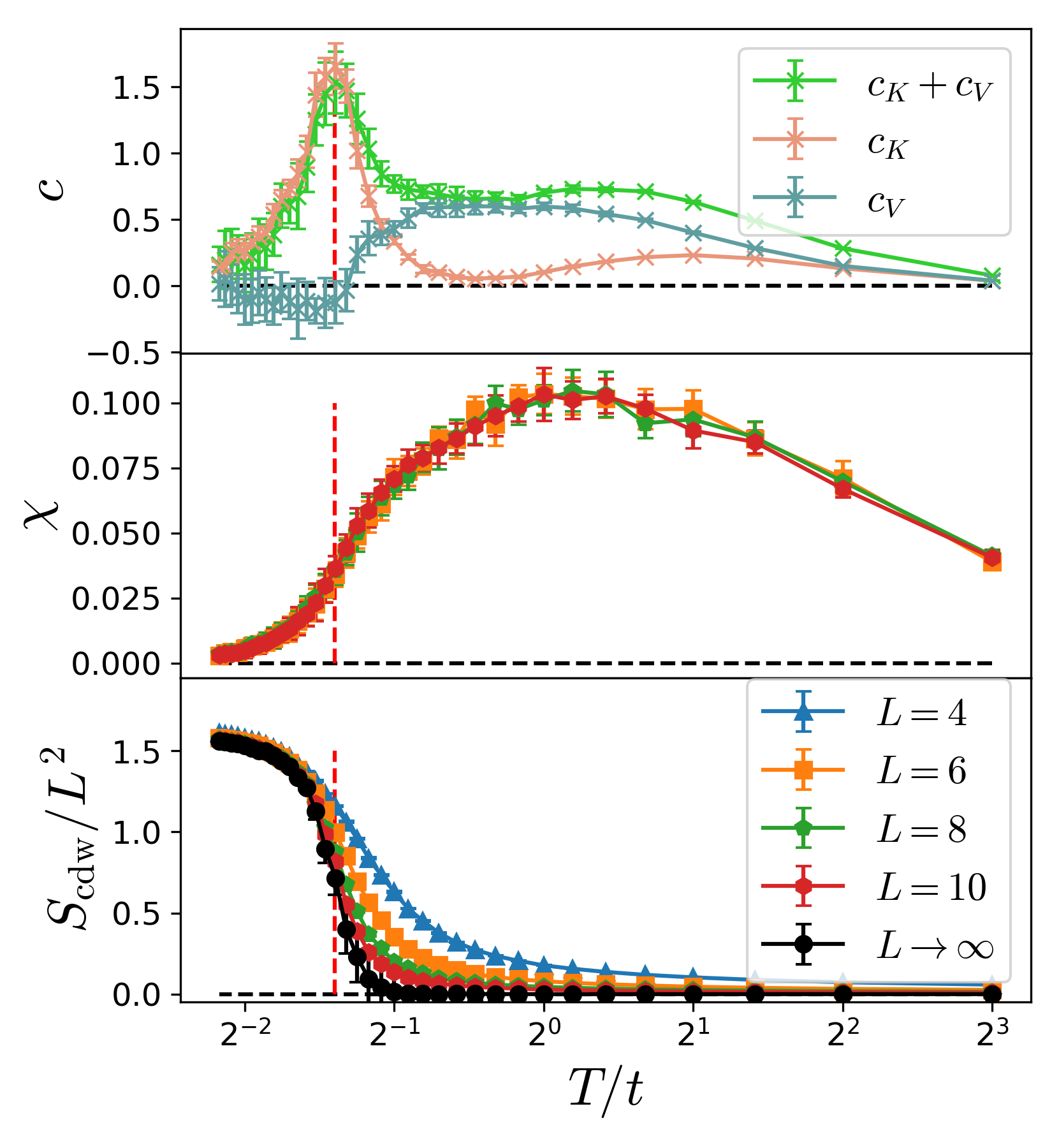}
   \caption{\label{fig:temp_cut} Heat capacity, difference susceptibility, and structure factor (finite-size extrapolation) as a function of \( T/t \) at \( U = 3.5, \mu = 0 \). The red dashed lines show the phase transition temperature \(T_c = 0.38 t\) as determined by the invariant correlation ratio method. Error bars in the $c$ and $\chi$ plots and $L = 4,6,8,10$ $S_{\rm cdw}/L^2$ data show statistical error, while error bars for the $L \to \infty$ $S_{\rm cdw}$ plot show the difference between the finite-size extrapolation and $L=10$ data.}
\end{figure}

We used 10,000 DQMC measurement sweeps for the data in Fig.~\ref{fig:phase_diagram} and 200,000 measurement sweeps for the data in Fig.~\ref{fig:mu_cut}.
In each of the above cases, $M = 96$.
For the data in Fig.~\ref{fig:temp_cut}, we used 200,000 measurement sweeps and varied $M$ such that \(t\Delta \tau \leq 1/24 \).
For each DQMC calculation, we used 5,000 warm-up sweeps before beginning measurements.

\textit{Results ---}
The main results of this Letter are shown in Fig.~\ref{fig:phase_diagram}.
At \( T = t/3 \), we observe three distinct regions in the \((U,\mu)\) plane:
a three-component Fermi liquid (FL) region, a trion liquid (TL) region, and an ordered CDW phase.
In order to distinguish these phases and study the transitions between them, we will first consider the trion formation process.
We measure the fraction of sites with each occupation number, which are given by
\begin{align}
  n^{(1)} &= \frac{1}{2L^2}\sum_{\substack{((\sigma,\tau,\nu)) \\ i}} n_{i\sigma} (1 - n_{i\tau}) (1 - n_{i\nu}),\\
  n^{(2)} &= \frac{1}{2L^2}\sum_{\substack{((\sigma,\tau,\nu)) \\ i}} n_{i\sigma} n_{i\tau} (1 - n_{i\nu}),\\
  n^{(3)} &= \frac{1}{6L^2}\sum_{\substack{((\sigma,\tau,\nu)) \\ i}}^{} n_{i\sigma}n_{i\tau}n_{i\nu},
\end{align}
where \( ((\sigma,\tau,\nu)) \) indexes the permutations of the three-element set of molecular spin states.

In the large-\( U \) limit, we expect to be in the TL region, where trions have formed and move coherently.
We use the difference susceptibility
\begin{align}
  \chi = \left(\pdv{\langle n_d\rangle }{\mu}\right)_{T, L} \label{eqn:chi_sus},
\end{align}
where \( n_d = n^{(2)} - n^{(1)} \), to detect this trion formation.
Our choice of this parameter requires some explanation.
In the large-\( U \) limit, trions are superpositions of the on-site trion state --- where molecules in each of the three spin states occupy a single site --- and off-site trion states --- where one of the molecules in the trion occupies an adjacent site \cite{pohlmann2013,xu2023}.
This virtual dissociation process persists even to zero temperature (as long as \( t/U > 0 \)) and causes an effective repulsion between trions since the process is suppressed when trions are on adjacent sites.
So, instead of using \( \expval{n^{(3)}} \) directly to indicate composite particles, we consider \( \expval{n_d} = \expval{n^{(2)} - n^{(1)}} \) which is the population difference between doubly- and singly-occupied sites.
This quantity will approach zero in the TL and CDW regions since the only population of doubly- and singly-occupied sites are those which are part of an off-site trion.
However at half-filling, \( \expval{n_d} = 0 \) due to the particle-hole symmetry, not the presence of trions.
So, we use \( \chi \) as our indicator since it is also zero in the TL and CDW regions while attaining a finite value throughout the FL region, even at half-filling.
A more detailed discussion of $\chi$ and a comparison with alternate trion detection methods \cite{chetcuti2023,xu2023}, may be found in Supplemental Material \cite{supp} Sec.~\suppChi.

As shown in Fig.~\ref{fig:phase_diagram} and Fig.~\ref{fig:mu_cut}, \( \chi \) attains a finite value when \( U \) is small, but approaches zero as \( U \) increases, across a wide range of chemical potential values.
This is consistent with our understanding of the TL region, as we expect trions to form at any finite density.
As the temperature is lowered, the crossover becomes sharper, seeming to evolve into a QPT as \( T \to 0 \).

To understand the regions on each side of this transition, we consider the isothermal compressibility,
\begin{align}
	\kappa = \frac{1}{\langle n\rangle ^2}\left(\pdv{\langle n \rangle }{\mu}\right)_{T,L} \label{eqn:kappa}.
\end{align}
As shown in Fig.~\ref{fig:mu_cut}, away from half-filling, e.g. for \( \mu = -0.5 t \), \( \kappa \) monotonically increases with \( U \).
In the large-\( U \) regime, we expect our system to behave as a weakly-interacting TL with an effective repulsion caused by the virtual dissociation process, which goes as \( t^2/U \) \cite{titvinidze2011}.
So, the numerical results qualitatively agree with this picture.

To better understand this behavior, we can compare \( \kappa \) to three limiting cases: a Hartree approximation for the weakly-interacting Fermi liquid, which should be exact when \( U/t = 0 \), a classical ideal gas of trions, which is valid for strong \( U \) and temperatures well above $t$, and the single-site atomic limit, which should be valid for strong $U \gg t$ (see Supplemental Material \cite{supp}, Secs. \suppComp.A, \suppComp.B and \suppComp.C).
We see that the Hartree approximation in Fig.~\ref{fig:mu_cut} matches our data very well and that while the classical ideal gas approximation is not perfect, it also qualitatively agrees with \( \kappa \) in the dilute case.
In particular, our data is much closer to the classical ideal gas of trions approximation than the single-site atomic limit, which is shown for \( T = t/2 \) in Fig.~\ref{fig:mu_cut}.
This provides evidence that the main contribution to the compressibility of the system is the motion of the trions, not the compressibility of the on-site trions themselves.

At half-filling (\(\mu/t = 0\)) and intermediate $U/t$, there is a drastic dip in $\kappa$, which seems to coincide with the trion formation QPT as indicated by $\chi$.
While this sort of compressibility minimum may be related to trion formation in similar models \cite{tajima2022,tajima2025}, the absence of this minimum away from half-filling leads us to suspect that some other physics is relevant.
At half-filling, an obvious candidate would be CDW ordering caused by the effective repulsion from the virtual dissociation process.
This should occur in some intermediate range of $U/t$, since $U/t$ must be large enough for trions to form, but the effective repulsion must also be strong enough to cause ordering.
This density ordering can be detected using the structure factor
\begin{align}
	S(\vb{q}) = \frac{1}{L^2}\sum_{i,j,\sigma,\tau}^{} e^{-i \vb{q}(\vb{r}_i - \vb{r}_j)} \langle n_{i\sigma} n_{j\sigma} \rangle \label{eqn:sq}.
\end{align}
We expect a two sublattice ``checkerboard'' order so we measure the \( \vb{q} = \vb{Q} \equiv (\pi,\pi) \) CDW structure factor, \(S_{\rm cdw} = S(\vb{Q})\), to use as an order parameter.

As shown Fig.~\ref{fig:phase_diagram}, this ordering indeed appears in a region (blue) near half-filling and at intermediate values of \( U \).
On the low-\( U \) side, trions are not able to form, so we do not expect ordering.
On the other hand, as \( U \) increases, the virtual dissociation becomes too weak to cause the ordering.
So the dip in \( \kappa\) near half-filling occurs because formation of CDW order is inhibiting the density response.
From the constant \( \mu \) cuts shown in Fig.~\ref{fig:mu_cut}, we can see that this ordering only appears sufficiently close to half-filling and at sufficiently low temperatures (it is absent for \( \mu \le -0.5 t \) or \( T \ge t/2 \)).

Based on the sharp behavior of \(S_{\rm cdw}\) near the transition, we conclude that this CDW region is bounded by a finite-temperature phase transition.
To obtain this sharp behavior, we use a finite-size extrapolation from $S_{\rm cdw}/L^2$ measured on different lattice sizes (see Supplemental Material \cite{supp} Sec. \suppFSS).
We also use the invariant correlation ratio to pinpoint this transition more precisely (details in Refs \cite{binder1981,kaul2015a} and Supplemental Material \cite{supp} Sec. \suppRcdw).
This method identifies a transition at both high and low \( U \) at half-filling along with a critical \( \mu/t \) value for \( U = 3.5 t\) .
This finite temperature phase transition contrasts with the $N=2$ case, where this ordering only appears in the ground state at half-filling and is not stable to finite $T$ \cite{scalettar1989,moreo1991,paiva2004}.
This is because the particle-hole symmetry is part of a continuous SU(2) symmetry in the $N=2$ case so the Mermin-Wagner theorem forbids finite-temperature ordering in two dimensions \cite{mermin1966}.
However, in the $N=3$ case at half-filling, the particle-hole symmetry is a discrete \( \mathbb{Z}_2 \) symmetry so the finite temperature ordering is allowed.
Further details of this argument are presented in Supplemental Material \cite{supp} Sec. \suppSym.

In order to further study the onset of CDW order and calculate a critical temperature \( T_c \), we computed \( \chi \) and \( S_{\rm cdw} \) along with the kinetic and potential energy components of the heat capacity \cite{paiva2001}
\begin{align}
    c_{\mathcal{O}} = \frac{1}{L^2}\left(\pdv{\mathcal{O}}{T}\right)_{\mu,L},
\end{align}
where $\mathcal{O}$ is $K$, $V$, or $K + V$.
The results of these calculations at \( U = 3.5t, \mu = 0 \) are shown in Fig.~\ref{fig:temp_cut}.

Here, the total heat capacity shows two peaks --- a narrow, tall peak which is almost entirely caused by the kinetic energy and a broader, shorter peak which has contributions from both components.
The sharp kinetic energy peak coincides with the onset of CDW order, as shown by the behavior of \( S_{\rm cdw} \), and the critical temperature calculated by the invariant correlation ratio method.
The behavior of \( \chi(T) \) near this transition temperature also shows that the trions form at similar temperatures to the CDW ordering, suggesting that the virtual dissociation process plays a role in stabilizing the trions near half-filling.

\textit{Discussion and Conclusion ---}
We have calculated properties of the SU(3) attractive FHM on a two-dimensional square lattice using the DQMC method.
We observed a finite-temperature CDW phase, as well as a FL-TL crossover at finite-temperature, which sharpens to a QPT at zero temperature.
These finite temperature results on a square lattice complement previous mean field \cite{honerkamp2004a,honerkamp2004}, dynamical mean field theory \cite{rapp2007,rapp2008,titvinidze2011} and self-energy functional \cite{inaba2009,inaba2011} studies along with quantum Monte Carlo results on a two-dimensional honeycomb lattice, restricted to half-filling \cite{xu2023,li2023,li2024}.

The QPT indicated by the FL-TL crossover occurs across a wide range of \( \mu \) away from half-filling and appears to occur at a finite $U_c > 0$.
This description is qualitatively similar to what we expect from the CSF to trion QPT for this model in higher dimensions, which has a critical $U_c > 0$ \cite{rapp2007,rapp2008}.
On the other hand, $U_c = 0$ for this model in one dimension \cite{lecheminant2005,capponi2007,capponi2008}, where CSF order only appears in the ground state in the presence of SU(3)-breaking interaction terms \cite{azaria2009}.
So, whether $U_c > 0$ in two dimensions is an open question.
We believe signatures of potential CSF order would be detectable in finite-$T$ DQMC calculations.
For a useful comparison, we consider the attractive SU(2) FHM on a square lattice.
While quasi-long-range superconducting order has been seen at finite temperature in the attractive SU(2) FHM on a two-dimensional square lattice \cite{moreo1991,paiva2004} and bilayer square lattice \cite{prasad2014,prasad2022}, it is separated from the normal phase by a Kosterlitz-Thouless transition \cite{kosterlitz1973}.
If a similar situation occurs for the attractive SU(3) case, we may be able to detect CSF order off half-filling (where the transition temperature may be finite \cite{scalettar1989}) and compare this transition to the FL-TL crossover \cite{inaba2009,inaba2011}).

Our results also complement previous results predicting the existence of CDW order at \( T = 0 \) \cite{honerkamp2004a,titvinidze2011}.
As noted before, in the SU(2) case this phase is not stable at finite temperatures.
So, our observation of both FL-CDW and TL-CDW finite temperature phase transitions suggests that this CDW phase is stable at finite temperature and opens up a number of questions around how this phase interacts with the trion formation crossover. 
Further studies into differences between the FL-CDW and TL-CDW transitions may lead to insights to the underlying physics, similarly to the different mechanisms which drive antiferromagnetism in the repulsive SU(2) FHM \cite{mott1949,slater1951,hirsch1985,paiva2011,chng2018}.
It is also not clear from our results what range the CDW order occupies in the ground state phase diagram.
One-loop RG methods predict that the ground state has long-range CDW order even for infinitesimal values of \( U \) in two-dimensions \cite{honerkamp2004a}, so we might expect CDW order at all values of \( U \) at a low enough temperature and half-filling.
In all of these questions --- whether the CSF phase exists in two dimensions, whether $U_c > 0$ for the FL-TL QPT, and what behavior of \( U_c \) is for the CDW transition near \( T=0 \) --- DQMC calculations performed at lower temperatures will be valuable for making progress.

Finally, we consider the experimental viability of observing these phases in an ultracold molecule optical lattice experiment.
The double microwave shielding technique \cite{karman2025}, which was used to suppress loss processes in the first dipolar BEC \cite{bigagli2024}, is believed to eliminate three-body loss in the form of three-body recombination processes so that only the two-body loss rate is relevant.
Considering only two-body loss processes and working in a strong optical lattice limit --- where we use the ground state simple harmonic oscillator as our Wannier function --- we compute that for \ce{NaCs} with $s$-wave scattering length \( a_s \approx 10^3 a_0 \), where $a_0$ is the Bohr radius, and two-body loss coefficient \( L_{\rm 2B} \approx \SI{3e-13}{\centi\meter^3/\second} \) (as reported in \cite{bigagli2024}), \( \hbar /U \tau_{\rm loss} \approx 10^3 \). 
The loss timescale is much longer than the timescale of the dynamics of our attractive SU(3) FHM implying double microwave shielding will suppress loss rates enough to reasonably simulate the attractive SU(3) FHM.

To measure signatures of the various phases discussed in this work, we propose using Quantum Gas Microscopy (QGM). 
QGM measures site-resolved snapshots of molecule locations, allowing for experimental access to \( \expval{n^{(1)}},\expval{n^{(2)}},\expval{n^{(3)}}\) and \( S_{\rm cdw}/L^2\).
These measurements can be directly compared to DQMC results for thermometry and consistency checks \cite{hart2015,cheuk2016,mazurenko2017}.
This technique has already been used to study the dynamics of interacting dipolar systems using ultracold molecules in an optical lattice \cite{christakis2023}. 
So, the recent development of microwave shielding for dipolar molecules should allow for flexible quantum simulation of a class of SU($N$) lattice models previously not experimentally accessible.
Continued theoretical investigation along with further experimental developments may yield exotic new phases and understanding of composite particles in many-body systems.

\textit{Acknowledgments ---}
K.R.A.H. and J.S. acknowledge support from the National Science Foundation (PHY-1848304)  and the W. M. Keck Foundation
(Grant No. 995764). 
RTS is supported by the grant DOE DE-SC0014671 funded by the U.S. Department of Energy, Office of Science.
This work was supported in part by the NOTS cluster operated by Rice University's Center for Research Computing (CRC).

\bigskip
\onecolumngrid
\newpage


\section*{Supplementary material}

# Supplemental Material for "Trion formation and ordering in the attractive SU(3) Fermi-Hubbard model"

Jonathan Stepp,[1, *] Eduardo Ibarra-García-Padilla,[2, 3] Richard T. Scalettar,[2] and Kaden R. A. Hazzard[1, 4]

[1]*Department of Physics and Astronomy, Rice University, Houston, Texas 77005, USA*
[2]*Department of Physics, University of California, Davis, California 95616, USA*
[3]*Department of Physics and Astronomy, San José State University, San José, California 95192, USA*
[4]*Smalley-Curl Institute, Houston, Texas 77005, USA*
(Dated: September 3, 2025)

## I. DQMC METHOD

We can write the Hamiltonian for the attractive SU($N$) Fermi-Hubbard model (FHM) as $H = Q + V$, where

$$Q = -t \sum_{\sigma,\langle i,j\rangle} \left[ c^\dagger_{i\sigma} c_{j\sigma} + \text{h.c.} \right] - \mu \sum_{\sigma,i} n_{i\sigma}, \quad (1)$$

$$V = -\frac{U}{2} \sum_i \left( \sum_\sigma n_{i\sigma} - \frac{N}{2} \right)^2. \quad (2)$$

Here, $i$ is the site index and $\sigma$ is the molecular spin index. Then, the Determinant Quantum Monte Carlo (DQMC) we have developed for the attractive SU($N$) FHM works by first Trotterizing the partition function

$$\mathcal{Z} = \text{Tr}\, e^{-\beta H} = \lim_{\Delta\tau \to 0} \left( e^{-\Delta\tau Q} e^{-\Delta\tau V} \right)^M, \quad (3)$$

where $\beta = \frac{1}{T}$, $T$ is the temperature (we set Boltzmann's constant $k_B = 1$ throughout), and $M \equiv \beta/\Delta\tau$ is the number of Trotter steps. We estimate the error caused by this Trotterization process and compare it to other sources of uncertainty in Section VI.

To handle the interaction term, $V$, we employ the Hubbard-Stratonovich transformation

$$\begin{aligned}
e^{-\Delta\tau V} &= \prod_i e^{\Delta\tau \frac{U}{2} \sum_i (n_{i\sigma} - \frac{N}{2})^2} \\
&= \prod_i \left( \frac{1}{\sqrt{2\pi}} \int_{-\infty}^\infty e^{-\frac{1}{2}x_i^2 - x_i\sqrt{\Delta\tau U}\left(\sum_\sigma n_{i\sigma} - \frac{N}{2}\right)} dx_i \right),
\end{aligned} \quad (4)$$

which decomposes the interaction term into a non-interacting term at the cost of introducing a classical field $x_i$ for each lattice site $i$. If we apply this Hubbard-Stratonovich transformation to each $e^{-\Delta\tau V}$ term in Eq. (3), we obtain

$$\mathcal{Z} = \left( \frac{1}{\sqrt{2\pi}} \right)^{ML^2} \int \prod_{\ell,i} dx_{i,\ell}\, \text{Tr} \left[ \prod_{\ell=1}^M e^{-\Delta\tau Q} e^{-\Delta\tau \tilde{V}_\ell} \right], \quad (5)$$

* jds28@rice.edu

where $\ell$ goes from 1 to $M$, $\tilde{V}_\ell = \sum_i -\frac{1}{2}x_{i\ell}^2 - x_{i\ell}\sqrt{\Delta\tau U}\left(\sum_\sigma n_{i\sigma} - \frac{N}{2}\right)$, and the classical fields $x_{i\ell}$ have gained an index to specify both their lattice site index, $i$, and imaginary time step index, $\ell$. Since $Q$ and each of the $\tilde{V}_\ell$'s are quadratic in fermion creation/annihilation operators, we can treat the integrand as a non-interacting system. If we calculate analytic results about this non-interacting system and then use a Monte-Carlo process to integrate over the $x_{i\ell}$ fields, we obtain results about the attractive SU($N$) FHM. As shown in Ref. [1], this results in

$$\begin{aligned}
\mathcal{Z} &= \left( \frac{1}{\sqrt{2\pi}} \right)^{ML^2} \int \prod_{\ell,i} dx_{i,\ell}\, \det M(\{x_{i\ell}\}) e^{-S_B(\{x_{i\ell}\})} \\
&= \left( \frac{1}{\sqrt{2\pi}} \right)^{ML^2} \int \prod_{\ell,i} dx_{i,\ell}\, [\det M^\sigma(\{x_{i\ell}\})]^N e^{-S_B(\{x_{i\ell}\})},
\end{aligned} \quad (6)$$

where $M(\{x_{i\ell}\})$ is a $NL^2 \times NL^2$ matrix and $S_B$ is a leftover bosonic action. Since there are no molecule spin state-changing interactions, $M(\{x_{i\ell}\})$ is block diagonal in the molecular spin states, and each block is identical due to the SU($N$) symmetry. Therefore, we can simplify the expression by defining $M^\sigma(\{x_{i\ell}\})$ as the $L^2 \times L^2$ matrix that makes up each block. Our notation emphasizes that each of these quantities depends on a specific configuration of the Hubbard-Stratonovich fields, $\{x_{i\ell}\}$.

During the Monte Carlo process, proposed changes are accepted based on the ratio between integrands in Eq. (6). So the probability of accepting a change $\{x_{i\ell}\} \to \{x'_{i\ell}\}$ is given by

$$p\left(\{x_{i\ell}\} \to \{x'_{i\ell}\}\right) = \frac{[\det M^\sigma(\{x'_{i\ell}\})]^N}{[\det M^\sigma(\{x_{i\ell}\})]^N} e^{-\Delta S_B}, \quad (7)$$

where $\Delta S_B = S_B(\{x_{i\ell}\}) - S_B(\{x'_{i\ell}\})$. Our DQMC implementation typically proposes changes which shift individual $x_{i\ell}$ variables by some amount one at a time. To improve convergence rates and avoid local minima, we also perform occasional "global moves" which propose changing every $x_{i\ell}$ for a given $i$ at once.

While this looks like a straightforward application of Monte-Carlo integration, the matrix $M^\sigma(\{x_{i\ell}\})$ is not necessarily positive-definite for every choice of $\{x_{i\ell}\}$, which means $p\left(\{x_{i\ell}\} \to \{x'_{i\ell}\}\right)$ is not always positive. When this happens, $|p\left(\{x_{i\ell}\} \to \{x'_{i\ell}\}\right)|$ can be used as the

acceptance probability, and the standard DQMC measurements can still be performed. However, if the positive and negative sign regions of configuration space have comparable weight, the number of DQMC sweeps needed to resolve observables to a given error grows exponentially with decreasing error.

This exponential increase in required computational resources is the so-called "fermion sign problem" [2–4]. Note that if $N$ is even, $[\det M^\sigma(\{x_{i\ell}\})]^N > 0$ for any field configuration so our DQMC implementation is sign problem free. The sign problem is negligible for the $N = 3$ results in this paper, but we expect it to become more significant as we lower the temperature and increase the system size.

## II. COMPRESSIBILITY APPROXIMATIONS

This section develops approximations to understand the small- and large-$U$ limits of the compressibility.

### A. Weakly-interacting mean field

We use a Hartree approximation to treat the attractive SU(3) FHM in the weakly interacting limit ($U \ll t, \mu$). First, we re-write our Hamiltonian in a more convenient form:

$$H = \sum_\sigma \left[ -t \sum_{\langle i,j \rangle} c^\dagger_{i\sigma} c_{j\sigma} + \text{h.c.} \right.$$
$$\left. - \sum_i \left( (\mu - U) n_{i\sigma} + \frac{U}{2} \sum_{\tau \neq \sigma} n_{i\sigma} n_{i\tau} \right) \right]. \quad (8)$$

We can then obtain our mean field Hamiltonian by expanding the density operators around their average values and dropping second-order terms. So,

$$H_{\text{MF}} = \sum_\sigma \left[ -t \sum_{\langle i,j \rangle} c^\dagger_{i\sigma} c_{j\sigma} + \text{h.c.} - \sum_i (\mu - U) n_{i\sigma} \right.$$
$$\left. - \frac{U}{2} \sum_{\tau \neq \sigma} \langle n_{i\sigma} \rangle n_{i\tau} + \langle n_{i\tau} \rangle n_{i\sigma} - \langle n_{i\sigma} \rangle \langle n_{i\tau} \rangle \right]. \quad (9)$$

Then, setting $\langle n_{i\sigma} \rangle = \bar{n}$ for each $i$ and $\sigma$, we obtain the non-interacting mean field Hamiltonian

$$H_{\text{MF}} = \sum_{\mathbf{k},\sigma} (\varepsilon_\mathbf{k} - \mu - U(2\bar{n} - 1)) n_{\mathbf{k},\sigma}$$
$$= \sum_{\mathbf{k},\sigma} (\varepsilon_\mathbf{k} - z) n_{\mathbf{k},\sigma}, \quad (10)$$

where $\mathbf{k} = (2\pi n_x/L, 2\pi n_y/L)$ for integers $n_x, n_y \in [0, L)$, $\varepsilon_\mathbf{k} = -2t(\cos(k_x) + \cos(k_y))$, and $z = \mu - U(2\bar{n} - 1)$.

The Hartree approximation is computed by self-consistently solving

$$\bar{n} = \frac{1}{L^2} \sum_\mathbf{k} \frac{1}{1 + \exp[\beta(\varepsilon_\mathbf{k} - z)]} \quad (11)$$

for $\bar{n}$. Then, the single-component isothermal compressibility is

$$\kappa = \frac{\kappa_0(\beta, t, z, L)}{1 - 2U\bar{n}^2 \kappa_0(\beta, t, z, L)}, \quad (12)$$

where

$$\kappa_0(\beta, t, z, L) = \frac{\beta}{\bar{n}^2 L^2} \sum_\mathbf{k} \frac{1}{2 + 2\cosh(\beta(\varepsilon_\mathbf{k} - z))} \quad (13)$$

is the isothermal compressibility for a non-interacting single fermion-species on an $L \times L$ square lattice.

### B. Classical ideal gas of trions

Starting from

$$\kappa = \frac{1}{n^2} \left( \frac{\partial n}{\partial \mu} \right)_{T,L}, \quad (14)$$

we can use the Gibbs-Duhem relation to write that

$$\kappa = -\frac{1}{V} \left( \frac{\partial V}{\partial p} \right)_{T,\mu}. \quad (15)$$

Then, using the equation of state for a classical ideal gas,

$$pV = NT, \quad (16)$$

we can compute

$$\kappa = \frac{1}{nT}, \quad (17)$$

where $n = \frac{N}{V}$ is the density of particles we are considering.

For our classical ideal gas approximation, we assume we are deep in the trion regime, so $n_{\text{tri}} = \frac{1}{3} \sum_\sigma n_\sigma$. For a useful comparison with our DQMC data, we calculate our $\kappa$ by using the density reported by our DQMC and Eq. (17).

### C. Atomic limit

For the atomic limit calculation, we solve the one site problem, which is a valid approximation in the $U \gg t$ limit. For a single site, we have that

$$\mathcal{Z} = \sum_{k=0}^{3} \binom{3}{k} e^{\beta \mu k + \frac{\beta U}{2}(\frac{3}{2} - k)^2}, \quad (18)$$

and

$$\langle n_\sigma \rangle = \frac{1}{\mathcal{Z}} \sum_{k=1}^{3} \binom{2}{k-1} e^{\beta\mu k + \frac{\beta U}{2}(\frac{3}{2}-k)^2}. \quad (19)$$

We also have that

$$\left(\frac{\partial \langle n_\sigma \rangle}{\partial \mu}\right)_T = \frac{\beta}{\mathcal{Z}} \sum_{k=1}^{3} k \binom{2}{k-1} e^{\beta\mu k + \frac{\beta U}{2}(\frac{3}{2}-k)^2}, \quad (20)$$

which allows us to calculate $\kappa$. Since we apply this approximation in the limit where $|\beta U| \gg |\beta\mu|$, we can obtain a useful limit by evaluating only $k = 0, 3$ terms in each of these sums. This gives us

$$\kappa = 3\beta(1 + e^{-3\beta\mu}), \quad (21)$$

which, for $\mu < 0$, increases exponentially with $\beta$.

## III. THE INVARIANT CORRELATION RATIO

As one way to pinpoint the finite-temperature phase transitions into the CDW phase, we introduce the invariant correlation ratio,

$$R_{\mathrm{cdw}} = 1 - \frac{S_{\mathrm{cdw}}(\boldsymbol{\Gamma} + \delta\mathbf{q})}{S_{\mathrm{cdw}}(\boldsymbol{\Gamma})}, \quad (22)$$

where $\boldsymbol{\Gamma} + \delta\mathbf{q}$ is a neighboring reciprocal lattice vector to $\boldsymbol{\Gamma} = (\pi, \pi)$. By finite-size scaling arguments, this quantity will reach 1 in an ordered state, 0 in a disordered state and achieves a lattice-size independent value at a critical point [5, 6]. This allows us to identify the transition point near the resolution of our data points without introducing an artificial cutoff value.

As shown in Fig. 1 crossing behavior occurs along the boundary of the CDW phase in the $U, \mu$ and $T$ directions which gives further evidence of a well-defined finite-temperature ordered phase. Here, the uncertainty is reported using the distance between successive data points.

## IV. NUMERICAL DIFFERENTIATION

For the calculation of $\chi$ and $\kappa$, we needed to take a derivative with respect to $\mu$. For both of these quantities, we used a simple first-order formula using only one additional DQMC run,

$$\frac{\partial \mathcal{O}}{\partial \mu} \approx \frac{\mathcal{O}(\mu) - \mathcal{O}(\mu - \Delta\mu)}{\Delta\mu}, \quad (23)$$

where $\mathcal{O}$ is the expectation value we are differentiating and $\Delta\mu = 0.05t$.

For heat capacity data, we chose points which were equally spaced in $\beta = 1/T$. So, we differentiated using the chain rule and a second-order derivative formula,

$$\frac{\partial \mathcal{O}}{\partial T} \approx \frac{1}{T^2} \frac{\mathcal{O}(\beta - \Delta\beta) - \mathcal{O}(\beta + \Delta\beta)}{2\Delta\beta}, \quad (24)$$

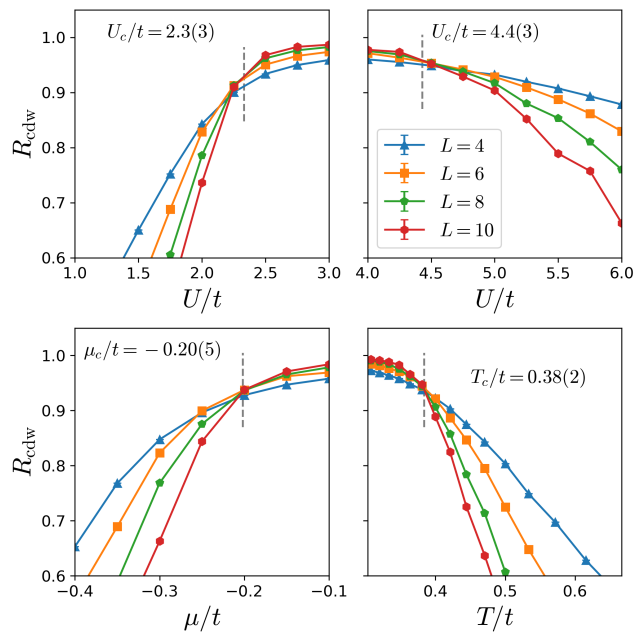

FIG. 1. The invariant correlation ratio for different lattice sizes. The crossings in the $U, \mu,$ and $T$ directions provide evidence of the CDW phase transition and estimate the phase boundary as a function of each variable. The $L = 8$ to $L = 10$ crossings are all indicated by dashed gray lines with the critical values listed in each plot. Plots are all at $\mu = 0, U = 3.5t, T = t/3$ unless otherwise specified by the axis label.

where $\mathcal{O}$ is the $T$-dependent quantity we are differentiating and where $t\Delta\tau = 0.125$.

In both cases, the statistical uncertainty reported for these quantities was calculated via error propagation of Eq. (23) and Eq. (24).

## V. FINITE-SIZE EXTRAPOLATION

Since the DQMC method calculates properties on a finite-size lattice and we are interested in phase transitions — which necessarily only occur in the infinite-size limit — it is necessary to extrapolate from the finite-size cases to the infinite-size case. The parameter most affected by this extrapolation is the CDW order parameter: $S_{\mathrm{cdw}}/L^2$. Near the CDW transition, $S_{\mathrm{cdw}}/L^2$ has a clear system-size dependence, so to extrapolate to the infinite-size case, we fit to the functional form

$$\frac{1}{L}\sqrt{S_{\mathrm{cdw}}\left(\frac{1}{L}\right)} = A + \frac{B}{L} + \frac{C}{L^2}. \quad (25)$$

This form has been used to analyze a corresponding CDW QPT for attractive, SU(3) fermions on a two-dimensional honeycomb lattice [7].

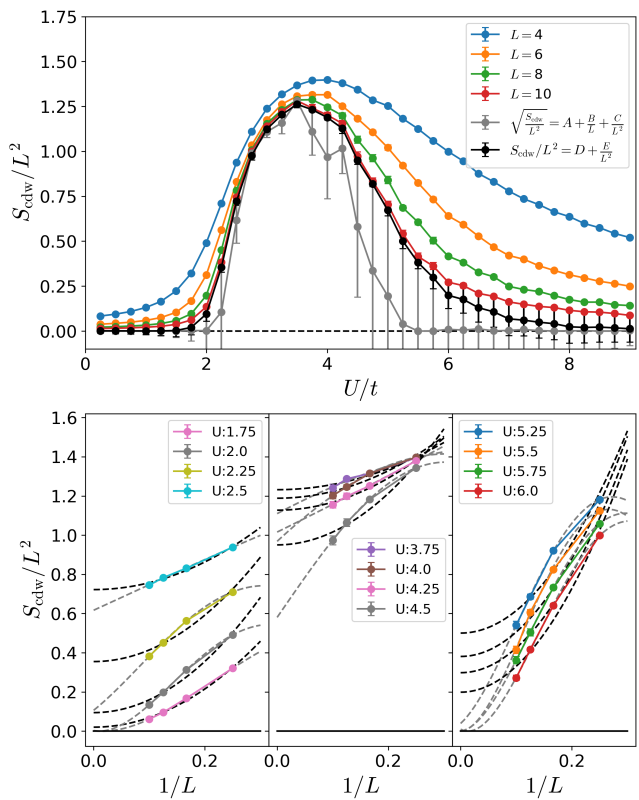

FIG. 2. Top: The CDW structure factor at $T = t/3$ and $\mu = 0$ for different lattice sizes. The finite-size extrapolations using Eq. (25) and Eq. (26) are shown in gray and black respectively. Bottom: The CDW structure factor as a function of $1/L$ along with the fits for various ranges of $U/t$.

On the other hand the functional form,

$$\frac{1}{L^2} S_{\text{cdw}}\left(\frac{1}{L}\right) = D + \frac{E}{L^2}, \qquad (26)$$

has also been used for a CDW transition on a honeycomb lattice and is motivated by spin wave theory [8].

We compare these two functional forms in Fig. 2. While the form in Eq. (26) seems smoother and closer to the $L = 10$ data point, we can see that the form in Eq. (25) resolves sharp behavior at both low- and high-$U/t$. Also, the form in Eq. (25) gives a high-$U/t$ transition out of the CDW phase which agrees more closely with the $U_c/t$ predicted by the invariant ratio analysis.

## VI. SOURCES OF UNCERTAINTY

The DQMC algorithm depends on a Trotter decomposition of the Hamiltonian as shown in Eq. (3). In Fig. 3, we compute the CDW structure factor for a range of $t\Delta\tau$ at a specific point in the CDW phase using the same number of DQMC sweeps as in other results. The smallest value of $t\Delta\tau$ in this chart is $t\Delta\tau = 1/24$ and we can see

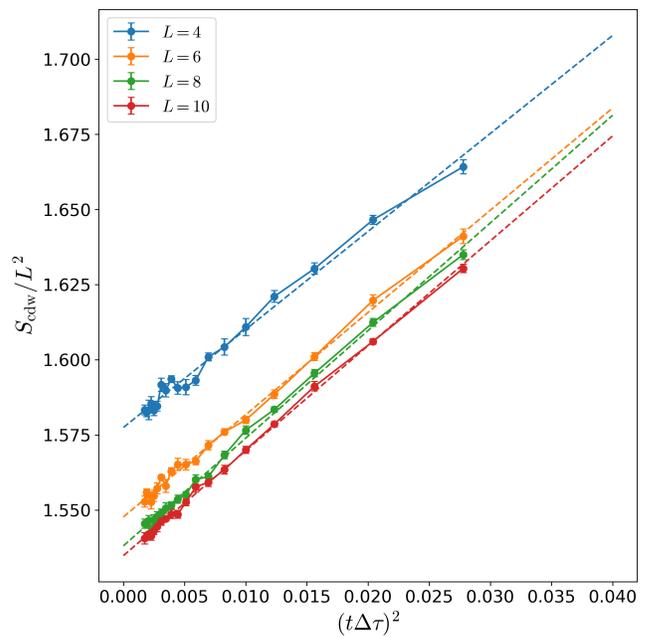

FIG. 3. The CDW structure factor at $U = 3.5t, \mu = 0, T = t/4$ as a function of $(t\Delta\tau)^2$. The dashed lines show a linear fit in $(t\Delta\tau)^2$.

that our results are well-converged at this value. It is also clear on this graph that we are able to resolve finite-size effects and observe that the Trotter error is of a similar magnitude to the DQMC statistical error. Additionally, we fit this data to a quadratic function in $t\Delta\tau$ and compute the relative error between our measured data point and the $t\Delta\tau \to 0$ extrapolation. For each size, this relative error is less than 1%.

|  | $L=4$ | $L=6$ | $L=8$ | $L=10$ |
|---|---|---|---|---|
| Trotter ($\times 10^{-3}$) | 3.659 | 3.266 | 4.759 | 3.740 |
| Finite-Size ($\times 10^{-3}$) | — | 19.623 | 4.751 | 3.144 |
| Statistical ($\times 10^{-3}$) | 1.035 | 1.190 | 0.993 | 1.211 |

TABLE I. Relative error of the CDW structure factor caused by Trotter error, finite-size effects and statistical error at $U = 3.5t, \mu = 0, T = t/4$ for each lattice size.

In Table I, we compare the relative error from each source at the $U = 3.5t, \mu = 0, T = t/4$ point for the different lattice sizes. The Trotter error is given by the difference between the last data point ($t\Delta\tau = 1/24$) and the $t\Delta\tau \to 0$ extrapolated value. The finite-size error is computed by the difference between the value for each lattice size and the value on the lattice one size smaller on the list. The statistical error is given by the standard deviation of the value as recorded by the DQMC algorithm. From these results, we see that these different sources of error are of a similar magnitude, with the statistical error generally being the smallest.


\newpage

\end{document}